\documentclass[twocolumn,showpacs,preprintnumbers,amsmath,amssymb,pra,superscriptaddress]{revtex4}
\usepackage{amssymb}
\usepackage{stmaryrd}
\usepackage{txfonts}
\usepackage{amsmath}
\usepackage{CJK}
\usepackage{graphicx}% Include figure files
\usepackage{dcolumn}% Align table columns on decimal point
\usepackage{bm}% bold math
\usepackage{ulem}
\usepackage{color}

\begin{document}
\begin{CJK*}{GB}{kai}
\CJKfamily{gbsn}

\preprint{}

\title{Coherent optical control of polarization with a critical
  metasurface}

\author{Ming~Kang}
\email{mingkang@mail.nankai.edu.cn}

\affiliation{College of Physics and Materials Science, Tianjin Normal University, Tianjin, 300387, China}

\author{Y.~D.~Chong}
\email{yidong@ntu.edu.sg}

\affiliation{School of Physical and Mathematical Sciences and Centre
  for Disruptive Photonic Technologies, Nanyang Technological University, Singapore 637371, Singapore}

\date{\today}

\begin{abstract}
We describe the mechanism by which a metamaterial surface can act as
an ideal phase-controlled rotatable linear polarizer.  With
equal-power linearly polarized beams incident on each side of the
surface, varying the relative phase rotates the polarization angles of
the output beams, while keeping the polarization exactly linear.  The
explanation is based on coupled-mode theory and the idea of coherent
perfect absorption into auxiliary polarization channels.  The
polarization-rotating behavior occurs at a critical point of the
coupled-mode theory, which can be associated with the exceptional
point of a parity-time (PT) symmetric effective Hamiltonian.
\end{abstract}


\maketitle
\end{CJK*}

In photonics, optical loss is commonly regarded as an unwanted
nuisance.  However, some recent advances have shown that loss is an
interesting control parameter in its own right, and can be used to
manipulate coherent light in useful ways.  A case in point is the
phenomenon of coherent perfect absorption (CPA): when the loss in an
optical structure is tuned to an appropriate (non-infinite) level, a
specific incident wavefront is absorbed without scattering
\cite{Chong10,Wan,Longhi11,Noh,Zhang,DuttaGupta,Pu,Gutman,Kang13a,Kang14,Zhang14}.
This is a generalization of the phenomenon of ``critical coupling''
\cite{Chong10}, and in a multi-channel system, like a metamaterial
surface (``metasurface'') with waves incident from both sides, it
provides a way to control light with light without optical
nonlinearity \cite{Zhang,DuttaGupta,Pu}: varying part of the incident
wavefront, such as the phase of one input beam, can switch the whole
wavefront between perfect and near-zero absorption.  Another example
of optical loss as a control parameter comes from the field of PT
symmetric optics, which deals with structures containing
spatially-balanced regions of gain and loss
\cite{Bender,Bender02,Moiseyev,Guo,Ruter,Bittner,Mostafazadeh,Longhi09,Longhi10,Chong11,Ge12,Lin,Regensburger,Castaldi,Peng}.
Such devices exhibit an unusual form of non-Hermitian
symmetry-breaking \cite{Bender}, and the ``critical point'' or
``exceptional point'' of PT symmetry-breaking \cite{Moiseyev} has been
found to be associated with extraordinary behaviors like
unidirectional invisibility
\cite{Guo,Ruter,Bittner,Mostafazadeh,Lin,Regensburger,Castaldi,Peng}.
Intriguingly, several links have been found between PT symmetry and
CPA.  PT symmetric scatterers can simultaneously exhibit CPA and
lasing \cite{Longhi09,Chong11,Ge12}, and in some metasurfaces the
occurrence of CPA can be mapped to the PT-breaking transition of an
effective Hamiltonian \cite{Kang13}.

The theory of CPA is agnostic about the nature of the loss
\cite{Chong10}, which can be some combination of Ohmic loss,
fluorescence---or even radiation into other coherent channels
\cite{crescimanno,stone13}.  At first glance, treating radiative loss
using the language of CPA may seem pointless, for the ``absorption''
of light from one input channel, and its complete transmission into
another channel, occurs in so simple a system as a non-scattering
waveguide.  However, when there are \textit{multiple} scattering
channels, the CPA concept can provide an interesting method for the
coherent manipulation of polarization \cite{Wang14}.

The control of polarization with pairs of coherent input beams has
been explored in recent experiments by Mousavi \textit{et
  al.}~\cite{Nikolay2014,Nikolay2015}, who showed that when two
equal-power linearly polarized beams are incident on an
appropriately-designed chiral metasurface (e.g.~an arrray of
asymmetrically-split wire rings), varying the relative phase $\phi$
beween the beams can induce a complete rotation of each output beam's
polarization angle, with the output ellipticity varying by $\lesssim
15 ^\circ$.  Hence, the metasurface functions as a phase-controlled
polarization rotator.  To explain this, Mousavi \textit{et al.}~noted
that, for a single input beam, the transmission is approximately
circularly polarized; say, left-circularly polarized (LCP).  To
explain the polarization rotation, the reflection of the other input
beam, incident from the opposite side, must be right-circularly
polarized (RCP).  This implies, by mirror symmetry, that for each
input beam the reflection and transmission have the \textit{same}
handedness.  This seems counter-intuitive, for a chiral resonance with
mirror symmetry along the propagation axis ought to emit to each side
with \textit{opposite} handedness.

In this paper, we present a theoretical study of an ideal two-sided
polarization-rotating metasurface, which reveals deep ties to the
concepts of CPA
\cite{Chong10,Wan,Longhi11,Noh,Zhang,DuttaGupta,Pu,Gutman,Kang13a,Kang14,Zhang14}
and PT symmetry
\cite{Bender,Bender02,Moiseyev,Guo,Ruter,Bittner,Mostafazadeh,Longhi09,Longhi10,Chong11,Ge12,Lin,Regensburger,Castaldi,Peng,Kang13}.
The metasurface, differing from Refs.~\cite{Nikolay2014,Nikolay2015},
contains pairs of coupled resonators radiating into different linear
polarization channels.  Using coupled-mode theory
\cite{R21,R22,R23,R24,R25,R26,R27}, we show that when linearly
polarized input beams are incident on each side, achieving perfect
conversion to the other polarization (i.e., $90^\circ$ rotation)
requires a specific balance between radiative and non-radiative loss
rates; this is analogous to ordinary CPA, which occurs at specific
intrinsic loss levels \cite{Chong10}.  For special ``critical''
choices of the frequency and loss parameters, the output beams become
exactly linearly polarized for all values of the relative input phase
$\phi$, with polarization angles varying with $\phi$.  Under one-sided
illumination, the reflected and transmitted beams have the same
handedness, which results from interference between direct and
indirect transmission processes.  However, the critical metasurface is
generally \textit{not} a perfect circular polarizer under one-sided
illumination, and the simple explanation involving phase-shifted RCP
and LCP components \cite{Nikolay2014,Nikolay2015} holds only in the
limit where the radiative and inter-resonator near-field coupling
rates dominate the dissipative loss rate, and the total loss is
negligible.  Nonetheless, when two input beams are applied, the
elliptically polarized reflection and transmission on each side
combine to give zero total ellipticity.  The critical points of the
coupled mode theory, where this behavior occurs, are the PT-breaking
points of a non-Hermitian effective Hamiltonian, whose eigenvalues
give the frequencies for the CPA-like perfect polarization conversion
condition. (Similar mappings to PT symmetric Hamiltonians have
previously been explored for CPA \cite{Kang13}, and for polarization
conversion under one-sided transmission \cite{Lawrence}.)  Using other
parameters choices, we can also switch the outputs between circular
and linear polarization by varying $\phi$.  We present full-wave
simulation results verifying the predictions of the coupled-mode
theory.

\begin{figure}
  \centerline{\includegraphics[width=0.46\textwidth]{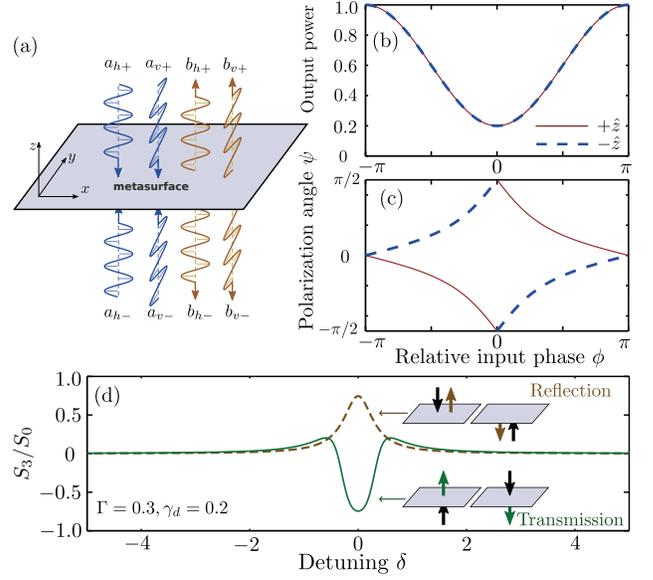}}
  \caption{(Color online) (a) Schematic of the input/output wave
    amplitudes $\vec{a}$ and $\vec{b}$, relative to the plane of the
    metasurface.  (b)--(c) Properties of the output beams emitted from
    the critical metasurface, versus relative phase $\phi \equiv
    \mathrm{arg}(a_{h-}/a_{h_+})$ of the equal-power input beams
    incident from each side.  The metasurface is tuned to the $\kappa
    = \Gamma = 0.3$, $\gamma_d = 0.2$ critical point.  (b) Output
    power $S_0 = |b_{h+}|^2 + |b_{h-}|^2$, with the power of each
    input beam normalized to 1, so that the total input power is 2.
    (c) Polarization angle $\psi = \tan^{-1}(S_2/S_1)/2$.  The outputs
    are linearly polarized ($S_3 = 0$) for all $\phi$.  (e)
    Ellipticity parameter $S_3/S_0$ versus detuning $\delta$ of a
    single input beam, under transmission from $\pm\hat{z}$ to
    $\mp\hat{z}$ (solid curve), and reflection from $\pm\hat{z}$ to
    $\pm\hat{z}$ (dashes).  At the critical point $\delta = 0$,
    neither reflected nor transmitted beam is exactly circularly
    polarized ($S_3/S_0 \ne 1$).}
  \label{fig:critical_output}
\end{figure}

Consider the plane metasurface depicted in
Fig.~\ref{fig:critical_output}(a), which is mirror-symmetric along the
direction $\hat{z}$ normal to the plane.  The metasurface is populated
by pairs of coupled resonant modes, described by amplitudes $\vec{q}
\equiv [q_x, q_y]^T$, which radiate in the $\hat{x}$ and $\hat{y}$
directions respectively.  Plane waves are normally incident on the
metasurface from both directions, with wave amplitudes $\vec{a} =
[a_{h+}, \,a_{h-}, \,a_{v+}, \,a_{v-}]^T$, where $+(-)$ denotes waves
incident from the top (bottom) of the plane, and $h (v)$ denotes the
linear polarization component parallel to the $\hat{x} (\hat{y})$
direction.  Likewise, the waves leaving the metasurface are described
by amplitudes $\vec{b} = [b_{h+}, \,b_{h-}, \,b_{v+}, \,b_{v-}]^T$,
with $\pm$ denoting waves exiting in the $\pm\hat{z}$ directions.  The
coupling between the metasurface resonances and the input/output waves
is described by a set of coupled-mode equations
\cite{R21,R22,R23,R24,R25,R26,R27}:
\begin{subequations}
\begin{align}
  -i\,\mathbf{\Omega}\, \vec{q} &= \mathbf{K}\, \vec{a} \label{Eq00}
   \\ \mathbf{K}^T\, \vec{q} + \mathbf{C}\,\vec{a} &= \vec{b},
   \label{Eq01}
\end{align}
\end{subequations}
where
\begin{subequations}
\begin{align}
  \mathbf{\Omega} &=
  \begin{pmatrix}
    \delta_x - i \left(\gamma_x^s +\gamma_x^d\right) & -\kappa \\ -\kappa &
    \delta_y - i \left(\gamma_y^s + \gamma_y^d\right) \\
  \end{pmatrix},  \label{Eq02} \\
  \mathbf{K} &= \begin{pmatrix}
    \sqrt{\gamma_x^s} & \sqrt{\gamma_x^s} & 0 & 0 \\
    0 & 0 & \sqrt{\gamma_y^s} & \sqrt{\gamma_y^s}
  \end{pmatrix},
  \label{Eq03} \\
  \mathbf{C}& = \begin{pmatrix}
    \sigma_1 & \mathbf{0} \\
    \mathbf{0} & \sigma_1
  \end{pmatrix}, \;\;\;\mathrm{where}\;\;
  \sigma_1 \equiv \begin{pmatrix}0 & 1 \\ 1 & 0
  \end{pmatrix}.
  \label{Eq04}
\end{align}
\end{subequations}
Here, $\delta_{\mu}$ (where $\mu\in\{x,y\}$) is the detuning of the
operating frequency from the $\mu$-oriented mode, $\gamma_{\mu}^s$ is
the radiative scattering rate, $\gamma_{\mu}^d$ is the non-radiative
dissipation rate, and $\kappa$ is the near-field coupling between the
modes; all these parameters are real.  The matrix $\mathbf{K}$
represents the radiative coupling between the metasurface and the
input/output waves, while $\mathbf{C}$ represents the direct coupling
between the waves.  The forms of these matrices are constrained by the
mirror symmetry of the metasurface, the definitions of energy and
power in terms of the coupled-mode quantities, and optical reciprocity
\cite{R22}.  The scattering matrix $\mathbf{S}$, defined by
$\mathbf{S} \, \vec{a} = \vec{b}$, takes the form
\begin{equation} \label{Eq05}
  \mathbf{S}=\mathbf{C}+i\,\mathbf{K}^T\mathbf{\Omega}^{-1}\mathbf{K}
  \equiv \begin{pmatrix}
    \mathbf{S}_{x} & \mathbf{D} \\
    \mathbf{D}& \mathbf{S}_{y}
  \end{pmatrix}.
\end{equation}
The sub-matrix $\mathbf{S}_x$ ($\mathbf{S}_y$) describes how light
incident in the $h(v)$ polarization scatters into the same
polarization, while $\mathbf{D}$ describes the cross-polarized
scattering.

We now assume that the metasurface is designed so the resonances have
the same dissipation rates and frequencies:
\begin{equation}
  \gamma_x^d = \gamma_y^d \equiv \gamma_d, \quad \delta_x = \delta_y
  \equiv \delta.
  \label{gamma_assumption}
\end{equation}
(The radiative scattering rates, however, can and will differ.)
Furthermore, we consider purely $h$-polarized incident illumination,
with input amplitudes $\vec{a}_h = [a_{h+}, a_{h-}]^T$.  The outputs
in each polarization are $\vec{b}_h = \mathbf{S}_x \, \vec{a}_h$ and
$\vec{b}_v = \mathbf{D} \, \vec{a}_h$.  Due to the mirror symmetry,
$\mathbf{S}_x$ has a symmetric eigenvector $[1,1]^T$ and an
antisymmetric eigenvector $[1,-1]^T$; the latter, with eigenvalue
$-1$, corresponds to a node on the plane, and hence zero total loss.

By varying the metasurface parameters and input amplitudes, it is
possible to arrive at a situation where $\vec{b}_h = 0$, i.e.~all the
$h$-polarized incident light is re-emitted in the $v$ polarization
and/or dissipated.  This corresponds to CPA with $v$-polarized
emission as one of the ``absorption'' channels (it would be
``ordinary'' CPA if $\gamma_y^s = 0$).  For this to occur, the
symmetric eigenvector of $\mathbf{S}_x$ must have eigenvalue zero, and
we can show that this occurs if and only if
\begin{subequations}
\begin{align}
  \gamma_x^s - \gamma_d &= \gamma_y^s + \gamma_d \; \equiv\; \Gamma,
  \label{gamma_condition} \\
  \delta^2 &= \kappa^2 - \Gamma^2. \label{delta_condition}
\end{align}
\end{subequations}
We can satisfy Eq.~(\ref{gamma_condition}) by designing each resonator
appropriately, as discussed below.  Then if $|\kappa| > \Gamma$,
perfect polarization conversion can occur at two distinct frequencies.
But if $|\kappa| < \Gamma$, Eq.~(\ref{delta_condition}) cannot be
satisfied for any $\delta$.

We can interpret the conditions
(\ref{gamma_condition})--(\ref{delta_condition}) in terms of an
effective Hamiltonian, via an argument from Ref.~\cite{Kang13}.  From
Eq.~(\ref{Eq05}), we write $\det(\mathbf{S}_x) =
\det\left(\mathbf{H}_{x} - \delta\cdot \mathbf{I} \right)/\,
         [-\det(\mathbf{\Omega})]$, where
\begin{equation}
  \mathbf{H}_{x} = \begin{pmatrix} i(\gamma_d-\gamma_x^s) & \kappa \\
    \kappa & i(\gamma_d + \gamma_y^s) \\
  \end{pmatrix}.
  \label{Eq06}
\end{equation}
The eigenvalues of $\mathbf{H}_{x}$ are the detunings for which
$\det(\mathbf{S}_x) = 0$.  These detunings should be real, but
$\mathbf{H}_x$ is non-Hermitian.  However, $\mathbf{H}_x$ becomes PT
symmetric \cite{Bender}, with $\mathbf{P} = \sigma_1$ and $\mathbf{T}$
the complex conjugation operation, when Eq.~(\ref{gamma_condition}) is
satisfied.  In that case, the eigenvalues of $\mathbf{H}_{x}$ are $\pm
\sqrt{\kappa^2 - \Gamma^2}$, which are the solutions to
Eq.~(\ref{delta_condition}); then the eigenvalues are real for the
PT-unbroken phase of $\mathbf{H}_x$, $|\kappa| > \Gamma$.  In the
PT-broken phase, $|\kappa |< \Gamma$, perfect polarization conversion
cannot occur for any real $\delta$.  At $\kappa = \pm\Gamma$, which
are the critical points of Eq.~(\ref{delta_condition}) and the
PT-breaking exceptional points \cite{Moiseyev} of $\mathbf{H}_x$,
perfect polarization conversion occurs at only one detuning, $\delta =
0$.

Now suppose the metasurface is tuned to one of the critical points,
satisfying Eqs.~(\ref{gamma_condition})--(\ref{delta_condition}) with
$\kappa = \Gamma$, $\delta = 0$.  For $h$-polarized inputs, the
coupled-mode equations give
\begin{align}
  b_{h\pm} &= \mp \frac{1}{2}\left(a_{h+} - a_{h-}\right) \\
  b_{v\pm} &=
  -\frac{i}{2}\sqrt{\frac{\Gamma-\gamma_d}{\Gamma+\gamma_d}}\left(a_{h+}
  + a_{h-}\right).
  \label{bvpm}
\end{align}
The outputs have equal power, and third Stokes parameters
\begin{equation}
    S_3^\pm \equiv -2\mathrm{Im} \left[b_{h\pm}b_{v\pm}^*\right]
    = \pm \sqrt{\frac{\Gamma-\gamma_d}{\Gamma+\gamma_d}} \;
    \frac{\left|a_{h+}\right|^2 - \left|a_{h-}\right|^2}{2}.
    \label{S3out}
\end{equation}
Hence, for equal-power inputs ($|a_{h+}|^2 = |a_{h-}|^2$), both
outputs are exactly linearly polarized ($S_3^\pm = 0$).

Part of this result is easy to understand: the outputs are
$v$-polarized for symmetric inputs (perfect polarization conversion),
and $h$-polarized for antisymmetric inputs (node on the plane).
However, Eq.~(\ref{S3out}) goes further, and states that the output
beams are linearly polarized for \textit{any} choice of input beam
phases.  Varying the relative phase rotates the output beams'
polarization angles between $[0,\pi/2]$.  This behavior is specific to
the critical metasurface.  (At the other critical point, $\kappa = -
\Gamma$, Eqs.~(\ref{bvpm}) and (\ref{S3out}) have opposite signs, and
the phase shift rotates the polarization in the opposite direction.
Thus, $\kappa \ne 0$ implies broken left-right symmetry on the
metasurface.)

The action of the critical metasurface as a rotatable linear polarizer
is shown in Fig.~\ref{fig:critical_output}(b)--(c).  At $\phi \equiv
\mathrm{arg}(a_{h-}/a_{h_+}) = 0$, perfect polarization conversion
occurs, and the power of each output beam reaches a minimum of
$(\Gamma-\gamma_d)/(\Gamma+\gamma_d)$ due to dissipation.  If the
dissipation is weak, $\gamma_d \ll \Gamma$, the total loss is $\approx
4\gamma_d / \Gamma$, and the polarization angle relative to the
$\hat{x}$ axis is $\psi \approx (\pi \pm \phi)/2$ for the outputs in
the $\pm\hat{z}$ direction.

To better understand the critical behavior, we examine the reflection
and transmission under one-sided $h$-polarized illumination.  As shown
in Fig.~\ref{fig:critical_output}(d), $S_3 > 0$ for reflection and
$S_3 < 0$ for transmission at the critical point, for either choice of
input direction.  From the definition of $S_3$, this means that a
single input beam produces reflected and transmitted beams with the
same handedness.  This is because the transmission is comprised of
direct transmission of the linearly-polarized input, and
re-emission from the metasurface resonators.  At the critical point,
the total transmission's handedness is opposite to the re-emitted
component, and the same as the reflected beam.
Fig.~\ref{fig:critical_output}(d) also shows that the transmission and
reflection are \textit{not} exactly circularly polarized for $\gamma_d
\ne 0$, but these elliptically polarized components are nonetheless
able to combine to form linearly polarized output beams.

\begin{figure}
\centerline{\includegraphics[width=0.46\textwidth]{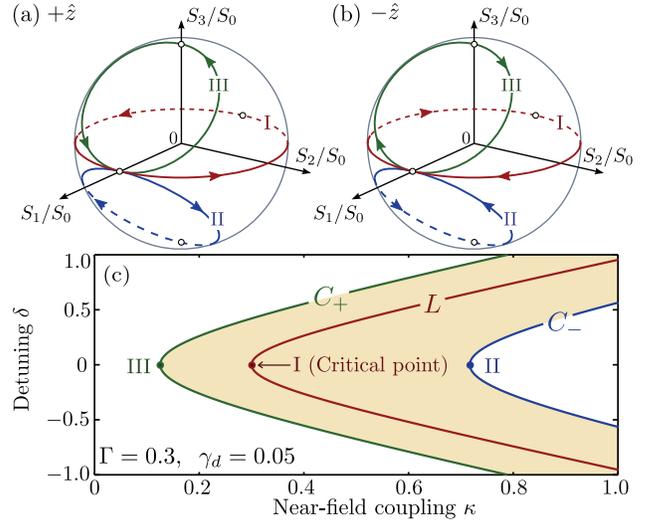}}
\caption{(Color online) (a),(b) Poincar\'e sphere trajectories of the
  output amplitudes for the (a) $+\hat{z}$ and (b) $-\hat{z}$ output
  beams, for equal-power input beams with relative input phase $\phi =
  \mathrm{arg}(a_{h-}/a_{h+})$.  Arrows indicate the direction with
  increasing $\phi$.  The trajectories labeled I, II, III have
  $\kappa, \,\delta$ parameters given by the matching points in (c);
  the other parameters are $\Gamma = 0.3$ and $\gamma_d = 0.05$.  (c)
  Phase diagram of the metasurface.  The curve labeled $L$ corresponds
  to the solutions of Eq.~(\ref{delta_condition}), for which
  $h$-polarized input beams can be perfectly converted to the $v$
  polarization; these are also the real eigenvalues of the PT
  symmetric effective Hamiltonian (\ref{Eq06}).  The curves labeled
  $C_+$ and $C_-$ correspond to solutions of Eq.~(\ref{Cline}), for
  which the inputs can be converted to circular polarization.  The
  $\kappa < 0$ part of the phase diagram is not shown here, but has a
  similar form.}
\label{fig:phasediag}
\end{figure}

Away from the critical point, the metasurface ceases to act as a
linear polarizer under equal-power incident beams, as each output beam
becomes elliptically polarized.  However, there is a remnant of the
critical behavior, due to the winding behavior on the Poincar\'e
sphere.  Right at the critical point, one cycle of $\phi$ winds each
output amplitude along the equator of the Poincar\'e sphere, as
depicted by the red loops in Fig.~\ref{fig:phasediag}(a)--(b).  Away
from the critical point, the trajectory no longer follows the equator
exactly, but one complete cycle of $\phi$ still induces one winding of
the longitudinal angle $2\psi$---and hence a full rotation of the
angle of the polarization ellipse's semi-major axis, $\psi$.  This is
true so long as the loops do not cross the poles (where the output
beams become circularly polarized).  It can be shown that the
pole-crossings occur when
\begin{equation}
  \delta^2 = \kappa^2 - \Gamma^2 \pm 2\kappa\sqrt{\Gamma^2 - \gamma_d^2}.
  \label{Cline}
\end{equation}
For $\kappa > 0$, the $+(-)$ signs correspond to an RCP (LCP)
$+\hat{z}$ output and an LCP (RCP) $-\hat{z}$ output.  This results in
the ``phase diagram'' shown in Fig.~\ref{fig:phasediag}(c).  The
solutions to Eq.~(\ref{Cline}) lie along the curves labeled $C_\pm$.
In the region between these two curves, the longitudinal angle $2\psi$
undergoes a complete winding with $\phi$.  The behavior for $\kappa <
0$ can be similarly deduced.  When Eq.~(\ref{Cline}) is satisfied,
varying $\phi$ switches the output between linear polarization and
circular polarization, as shown by the green and blue curves in
Fig.~\ref{fig:phasediag}(a)--(b).

\begin{figure}
\centerline{\includegraphics[width=0.9\columnwidth]{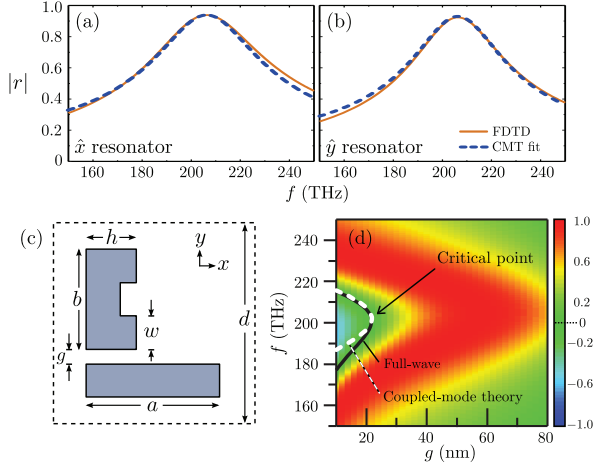}}
\caption{(Color online) Design of the critical metasurface.  (a)
  Reflectance spectrum for a strip antenna of length $a=434.25$~nm,
  width $w=50$~nm, and thickness $t=30$~nm, with incident beam
  linearly polarized parallel to the strip.  (b) Reflectance spectrum
  for a split-ring resonator (SRR) of width $w=50$ nm, main axis
  length $b=355$ nm, arm length $h=105$ nm, and thickness $t=30$ nm,
  with incident light linearly polarized parallel to the main axis.
  Both antennas are free-standing; solid lines show full-wave
  simulation results, while dashed lines show the least-squares fit to
  coupled-mode predictions.  (c) Schematic of the metasurface unit
  cell, with the strip antenna aligned in $\hat{x}$ and the SRR
  aligned in $\hat{y}$, separated by distance $g$. (d) Full-wave
  simulation results for $S_3/S_0$, versus $g$ and operating frequency
  $f$, for the $+\hat{z}$ output beam with symmetric equal-power input
  beams.  The dashed line shows where the coupled-mode condition
  Eq.~(\ref{delta_condition}) is satisfied, using fitted coupled-mode
  parameters; the solid black line shows where $S_3 = 0$ in the
  simulation results.  }
\label{fig:cmtfit}
\end{figure}

\begin{figure}
\centerline{\includegraphics[width=0.95\columnwidth]{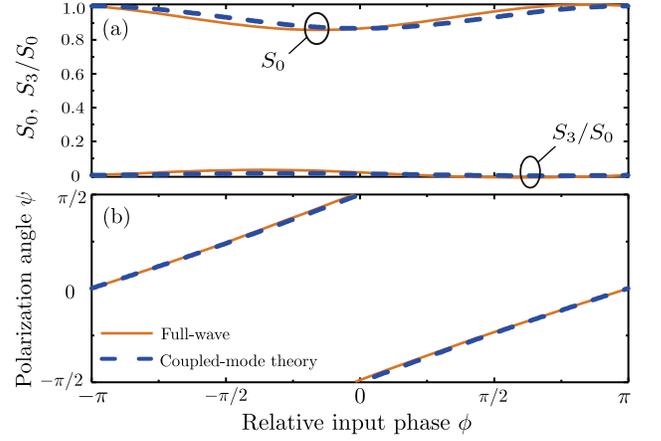}}
\caption{(Color online) Intensity $S_0$, ellipticity parameter
  $S_3/S_0$, and polarization angle $\psi$ for the $+\hat{z}$ output
  beam, versus the relative phase $\phi$ of the input beams.  The
  metasurface is at the critical point indicated in
  Fig.~\ref{fig:cmtfit}(d).  Solid curves show full-wave simulation
  results, and dashes show coupled-mode predictions using best-fit
  parameters. }
\label{fig:critical polarization}
\end{figure}

We have verified the above coupled-mode theory using numerical
simulations of an exemplary plasmonic metasurface, shown schematically
in Fig.~\ref{fig:cmtfit}(c).  Each unit cell contains a silver strip
antenna radiating in the $\hat{x}$ direction, and a silver split-ring
resonator (SRR) radiating in $\hat{y}$. The cells are arranged in a
square lattice with period $d=600$~nm, and the entire metasurface is
free-standing in vacuum, which guarantees that no high-order
diffraction mode exists below $500$~THz.  The dielectric function of
silver is modeled by a Drude formula $\varepsilon_m =
\varepsilon_{\infty} - f_p^2 / (f^{2} + i \gamma_p f)$, where $f_p =
2230$~THz, $\gamma_p = 5.09$~THz, and $\varepsilon_{\infty} = 5 $. To
extract the coupled-mode parameters for each resonator, we perform
full-wave (finite-difference time-domain) numerical simulations of
single-sided illumination incident on each antenna separately, in the
absence of the other antenna, with the appropriate linear
polarization.  The computed reflectance spectrum is fitted to the
theoretical result $R =
\gamma^{s\,2}_{\mu}/\left[(f-f_\mu)^2+(\gamma_\mu^s-\gamma_\mu^d)^2\right]$
obtained from Eqs.~(\ref{Eq01})--(\ref{Eq04}) in the $\kappa = 0$
limit.  Using the parameters stated in the caption of
Fig.~\ref{fig:cmtfit}, the resonators have equal resonance frequencies
$f_x= f_y = 206.3$ THz and dissipation rates $\gamma^{d}_{x} =
\gamma^{d}_{y}=1.3$~THz, thus satisfying Eq.~(\ref{gamma_assumption}).
Furthermore, the radiative decay rates are $\gamma^{s}_{x}=19.8$ THz
and $\gamma^{s}_{y}=17.2$ THz, satisfying Eq.~(\ref{gamma_condition}).

We then include both resonators in the metasurface, separated by
distance $g$, and perform another set of full-wave simulations.
Varying $g$, with all other geometrical parameters fixed, alters the
near-field coupling parameter $\kappa$, as well as (weakly) the
resonant frequency $f_0$.  Nonlinear fits of the simulation results to
the coupled-mode theory give the functional relations
$\kappa\approx28.32 -0.51g + 0.0026g^2$, where $\kappa$ and $g$ have
units of THz and nm, and $f_{0}\approx200.48+0.073 g$ in THz.  The
fitted coupled-mode theory gives results for the various output Stokes
parameters that agree well with the simulation results.  For instance,
Fig.~\ref{fig:cmtfit}(d) shows the simulation results of $S_3/S_0$ for
one of the output beams, with a pair of symmetric input beams.  The
locus of $S_3 = 0$ according to the simulation (black line) closely
matches coupled-mode prediction (white dashes).  Using
Eq.~(\ref{delta_condition}), we find that the critical point occurs at
$g \approx 22$ nm, $f \approx 202$ THz.  Fig.~\ref{fig:critical
  polarization} shows the behavior at this critical point, which is in
good agreement with the coupled-mode theory, particularly in the fact
that $S_3 \approx 0$ for all $\phi$.

In summary, we have shown that the principles of CPA and PT symmetry
can be used to design metasurfaces for coherently manipulating
polarization using two input beams.  This is an example of an emerging
class of photonic devices that exploit the properties of ``critical
points'' or ``exceptional points'' \cite{Lin,Peng,Rotter12,Rotter14}.
Here, the critical behavior corresponds to the output beams achieving
exactly zero ellipticity.  With other choices of metasurface
parameters, other forms of polarization control can be achieved, such
as switching between circular and linear polarization.
Generalizations of the coupled-mode theory, such as having resonators
which are not exclusively coupled to a single linear polarization
channel, might be useful for designing other classes of
polarization-controlling metasurfaces.

M.~Kang acknowledges supports from the National Natural Science
Foundation of China under Grant 11304226, and the China Postdoctoral
Science Foundation under Grant No.~2014M560414.  Y.~D.~Chong
acknowledges support from the Singapore National Research Foundation
under Grant No.~NRFF2012-02, and from the Singapore MOE Academic
Research Fund Tier 3 Grant MOE2011-T3-1-005.  We are grateful to
N.~I.~Zheludev, R.~Singh, C.~Altuzarra, and S.~Vezzoli for helpful
discussions and comments.

\end{document}